\documentclass[aps,pre,amsmath,amssymb,amsbsy,twocolumn,superscriptaddress]{revtex4-1}
\usepackage{amsfonts} 
\usepackage{amsmath}
\usepackage{amssymb}
\usepackage{cancel} 
\usepackage{bm}
\usepackage[usenames, dvipsnames]{color} 
\usepackage{dcolumn}
\usepackage{epic} 
\usepackage{epsfig}
\usepackage{eucal} 
\usepackage{graphicx}
\usepackage[breaklinks,colorlinks = true,linkcolor = red,urlcolor=blue,citecolor=red]{hyperref}
\usepackage{mathrsfs}
\usepackage{soul}
\usepackage{subfigure}
\usepackage{wrapfig} 
\usepackage{xy} 
\begin{document}
%%%%%%%%%%%%%%%%%%%%%%%%%%% 

%%%%%% TITLE %%%%%%%%%%%%%%
\title{Effect of heavy impurities on the dynamics of supercooled liquids}

%%%%%%%%%%%%%%%%%%%%%%%%%%% 

%%%%%% AUTHORS %%%%%%%%%%%%
\author{Saurish Chakrabarty}
% \email{schakrab@go.wustl.edu}
\affiliation{School of Chemical and Biomedical Engineering, Nanyang Technological University, Singapore 637459}
\affiliation{Physics Department, Acharya Prafulla Chandra College,
  Kolkata 700131, India}
\author{Ran Ni}
\email{r.ni@ntu.edu.sg}
\affiliation{School of Chemical and Biomedical Engineering, Nanyang Technological University, Singapore 637459}
%%%%%%%%%%%%%%%%%%%%%%%%%%% 

%%%%%%%% ABSTRACT %%%%%%%%%
\begin{abstract}
  We study the effect of heavy impurities on the dynamics of supercooled
  liquids. When a small fraction of particles in the supercooled
  liquid is made heavier, they exhibit slower dynamics than the original
  particles and also make the overall system slower. If one looks at
  the overlap correlation function to quantify dynamics in the system,
  it has different behavior for the heavy and the light particles. In
  particular, at the relaxation time of the overall system, the degree
  of relaxation achieved by the heavier particles is lesser on average
  than that achieved by the light particles. This difference in
  relaxation however, goes down drastically as a crossover
  temperature, $T_0$, is crossed. Below this crossover temperature,
  particles in the system have similar relaxation times irrespective
  of their masses. This crossover temperature depends on the fraction
  of the heavy particles and their masses. 
  Next, we isolate the effect of mass heterogeneity on the dynamics of
  supercooled liquids and find that its effect increases monotonically
  with temperature.
  We also see that the development of dynamical heterogeneity with
  decreasing temperature is less dramatic for the system with
  impurities than for the pure system. 
  Finally, the introduction of heavy impurities can be seen as a way of reducing the kinetic fragility of a
  supercooled liquid.
\end{abstract}
\date{\today}
%%%%%%%%%%%%%%%%%%%%%%%%%%% 
\maketitle
%%%%%%%%%%%%%%%%%%%%%%%%%%% 
\section{Introduction} 
The role of growing length-scales in triggering the glass transition
in supercooled liquids has been debated for the last few
decades.\cite{RFOT, cdggrowing, 10CG, 11keysPRX, WyartCatesNoLength, bbbt18length}
Even
though there is some consensus about the existence of a growing
dynamic length-scale and the phenomenon of dynamical heterogeneity,
its importance in the glass transition 
phenomenon is not yet clear. On the other hand, 
the structural signature correlated with the growing dynamic length-scale remains controversial~\cite{tanaka2008,tanaka2012,tanaka2013,tanaka2015,tanaka2018,small2019,small2020,liu2016,tanaka2019}, and 
the very existence of a
universal static length-scale across all glassy systems is still
questioned.\cite{10CG, 11keysPRX}
It is therefore important to focus on each of these length-scales and
understand what causes them to grow and how to suppress or enhance
their growth. This may help in
establishing whether the glass transition is indeed a result of the
divergence of one or both of them.
In this Letter, we try to control the amount of dynamical
heterogeneity present in a supercooled liquid by manipulating the
masses of the constituent particles.
We present a situation in which we alter the growth of
the dynamic length-scale without affecting its
static properties.
% We also present a situation where an imposed
% dynamical heterogeneity becomes less effective with decreasing
% temperature. 

We take a supercooled liquid and study its dynamics
with varying temperature. We then make a new system as follows. We
choose a small fraction of particles in the liquid and increase their mass
by a large constant factor.
Even though this step is similar in spirit to the idea of random
pinning, the important difference is that no disorder averaging is
needed in this case to match the thermodynamic properties of the
system with impurities with the pure system.
All particles in the
new system still enjoy the same degrees of freedom as they did in the
pure system, while randomly pinning the position of a few particles changes
the number of degrees of freedom in the system.
The potential energy landscape of the system is also left
unchanged. 

The main findings of this Letter are as follows.
First, the dynamic heterogeneities in the system with mass impurity and the pure system
are found qualitatively different, and in the system with mass impurity
the difference between the relaxation dynamics of the heavy
and light particles jumps at a
crossover temperature, $T_0$,  which is a function of the fraction of
heavy particles and their masses.
Second, if we factor out the effect of mass increase and isolate the
effect of mass heterogeneity alone, we see that the system with the
heterogeneity in masses is faster than the pure system and this
difference monotonically increases with temperature.
Third, the presence of heavy impurities results in a slowdown of the
growth of the dynamic length-scale from its high temperature value.
Fourth, there is an apparent decrease in the kinetic fragility because
of the presence of heavy impurities. 
\begin{figure}
  \centering
  \hspace{-.15in}\includegraphics[width=.9\columnwidth]{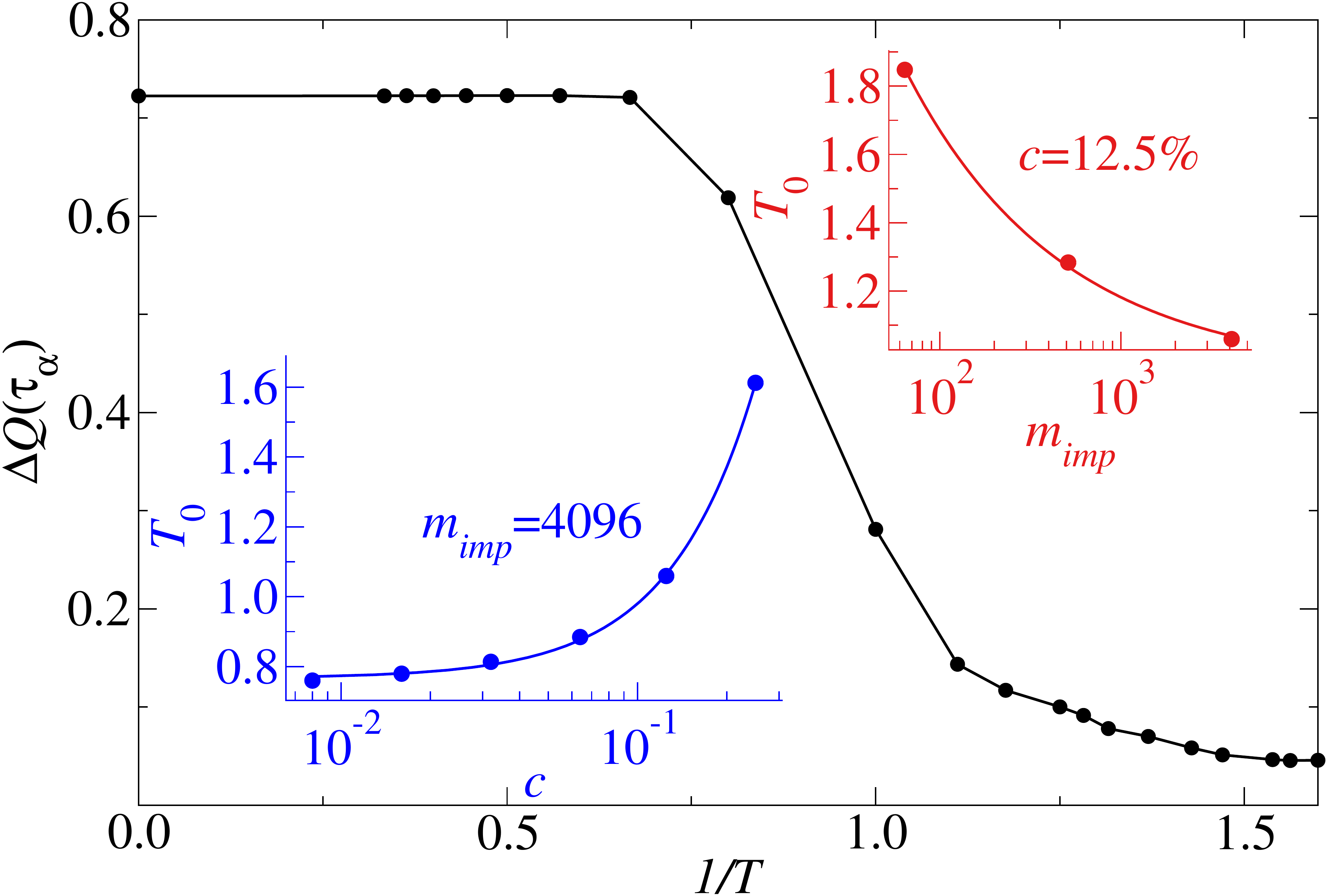}
  \caption{(Color online) 
    $\Delta Q(\tau_\alpha)$ plotted against inverse
    temperature. The infinite temperature point is obtained
    analytically. The value of $\Delta Q(\tau_\alpha)$ shows a significant
    drop across some temperature $T_0$ that is dependent on the mass of
    the heavy particles and the fraction of them present. In the top
    inset, we see that $T_0$ decreases with $m_{imp}$ and in the bottom
    one we see that it increases with the fraction of heavy
    impurities. (The curves in the insets are fits to the form shown
    in Eq. \ref{t0cm}.)
  }
  \label{deltaQ}
\end{figure}
\begin{figure}[h]
  \vspace{-.1in}
  \centering
  \includegraphics[width=.9\columnwidth]{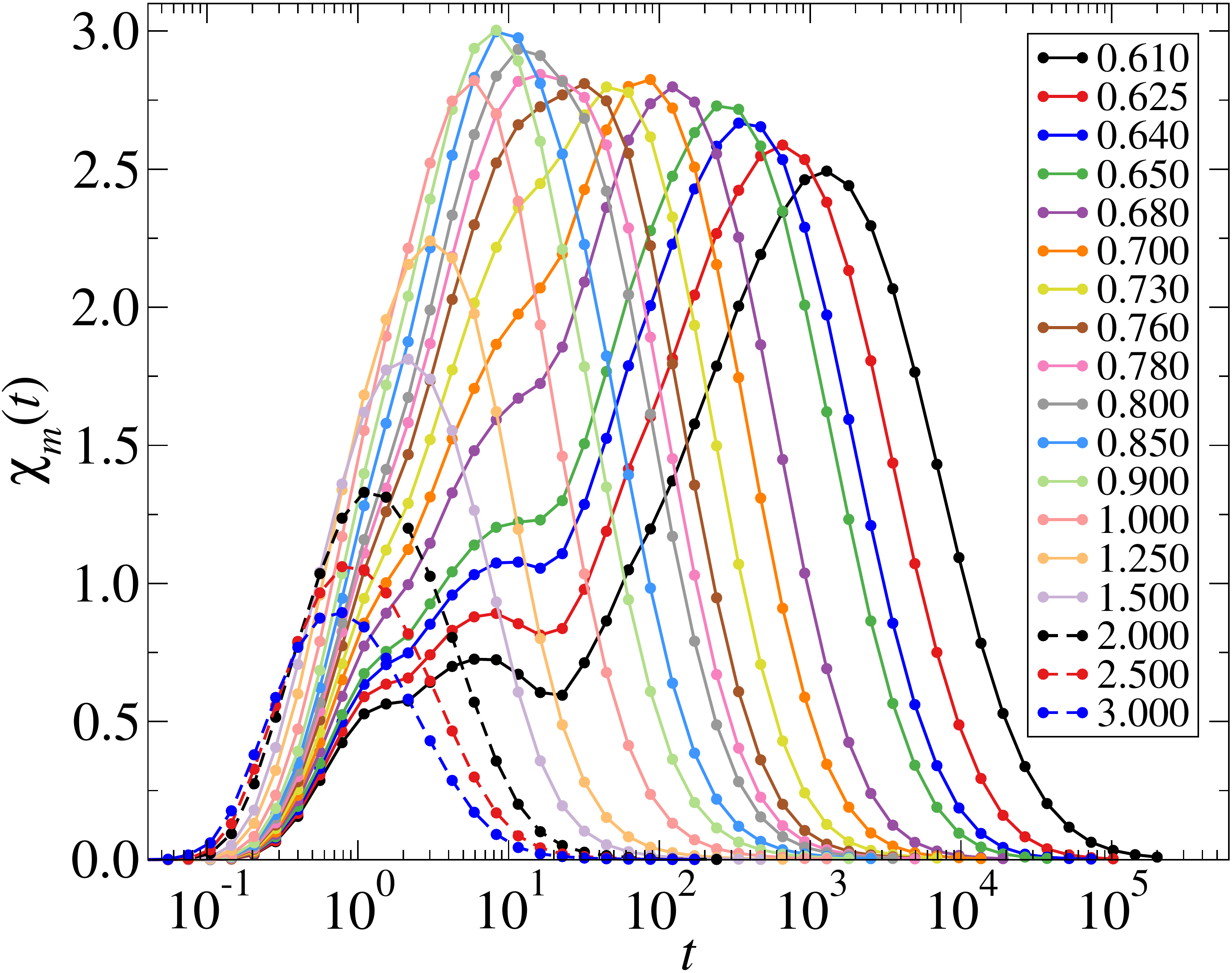}
  \caption{(Color online) 
    $\chi_m(t)$
    versus time for various temperatures in a system with $c=12.5\%$ and
    $m_{imp}=4096$.
  }
  \label{chimFig}
\end{figure}

\section{System studied}
We study a polydisperse system of Lennard-Jones particles with mean
diameter $\bar\sigma=1$ and having a standard deviation of $\Delta=12\%$. 
\begin{eqnarray}
  V_{ij}=4\epsilon\left[
  \left(\frac{\sigma_{ij}}{r_{ij}}\right)^{12}
  -
  \left(\frac{\sigma_{ij}}{r_{ij}}\right)^6
  \right],\\
  r_{ij}=|\vec{r}_i-\vec{r}_j|,~\epsilon=1 ~~~~~~~~~~\nonumber
\end{eqnarray}
The potential was truncated and shifted to zero at $r_{cut}=2.5\sigma_{ij}$.
The polydispersity in this system is discrete. There are $n_{poly}=100$
types of particles in equal proportion. The effective ``diameters'' take $n_{poly}$
equally spaced values in the range
$\left[\bar\sigma-\Delta\sqrt{\frac{3(n_{poly}-1)}{n_{poly}+1}},
  \bar\sigma+\Delta\sqrt{\frac{3(n_{poly}-1)}{n_{poly}+1}}\right]$. For a pair of particles of
different types, the effective $\sigma$ parameter is taken as the
average of the two particle diameters.
All particles in the system have unit mass. 
The density of the system is chosen so that the ``packing fraction''
defined as in Ref. \cite{10takeshiTanaka} is
$\phi=\frac{\pi}{6V}\sum_i\sigma_{ii}^3=0.55$. 
The system with impurities is obtained by choosing a fraction $c$ of
particles in the original system and changing their mass to
$m_{imp}>1$. Most of the results presented are for a system of
$N=1000$ particles. However, to calculate the dynamic length-scale we
used a system of $N=8000$ particles.

The temperature was measured in the unit of $\epsilon$ and time
in units of $\bar{\sigma}\sqrt{m/\epsilon}$ with $m$ is the
mass of a particle in the pure system equal to unity.

The systems were evolved in an $NVT$ ensemble, with temperature being
held fixed using the Nose-Hoover thermostat. The LAMMPS 
package was used for doing the simulations.\cite{lammps1,lammps2}
The dynamics was studied using the ensemble averaged
overlap correlation function, $Q(t)$,
which is the fraction of particles in the system which have moved by
$w=0.3\bar\sigma$ or less in time $t$.
When comparing the overlap correlation functions of the pure system
and the system with impurities, we use $Q(t)$ to denote the overlap
function for the pure system and $Q(c,m_{imp},t)$ to denote the
overlap function for a system in which a fraction $c$ of the particles
have been made heavier by a factor $m_{imp}$, $i.e.$, $Q(c,m_{imp},t)$
is the fraction of particles (heavy or light) that have moved by less
than $w$ in time $t$ in the system with impurities. In addition, in
the system with impurites, we
use $Q^{(m)}$ to denote the overlap function for particles of mass
$m$, $i.e.$, the fraction of particles of mass $m$ that have moved by
less than $w$ in time $t$. Thus,
$Q(c,m_{imp},t)=cQ^{(m_{imp})}(t)-(1-c)Q^{(1)}(t)$.

\section{The crossover in relaxational dynamics and the crossover
  temperature} We calculate the overlap 
correlation functions for the heavy and light particles separately in
a system with a fraction $c$ of heavy impurities with mass $m_{imp}$. 
Their difference, $\Delta Q(t)$, measures the difference in the
relaxation of the heavy and light particles.
\begin{eqnarray}
  \Delta Q(t)=Q^{(m_{imp})}(t)-Q^{(1)}(t),
\end{eqnarray}
where $Q^{(m_{imp})}$ and $Q^{(1)}$ are the overlap correlation functions
for heavy and light particles, respectively.
We focus on the value of
this quantity at the relaxation time, $\tau_\alpha$ of the overall
system [$Q(c,m_{imp},\tau_\alpha)\equiv1/e$]. This quantity, $\Delta
Q(\tau_\alpha)$, shows a jump at some 
crossover temperature $T_0$, that depends on $c$ and $m_{imp}$ as
shown in Fig. \ref{deltaQ}. 
% The temperature, $T_0$, can be
% thought of as a temperature below which the dynamics of the light
% particles are influenced significantly by the presence of the slow
% heavy particles, and vice versa.
The temperature, $T_0$, can be thought of as a temperature below which
the overall dynamics of the liquid is governed by collective phenomena
and the masses of the individual particles have little role to play.
\cite{isotopeMetallicGlass}
For a fixed fraction of heavy particles, this temperature
decreases with increasing mass of the heavy particles. For a fixed
mass of the heavy particles, it increases with the fraction of them
present. The asymptotic high temperature value of $\Delta
Q(\tau_\alpha)$ can be calculated analytically. In the high
temperature limit, it is expected that most of the light particles
move by a distance more than $w$ before any of the heavy particles
move by $w$. Thus, we get, 
\begin{eqnarray}
  Q_\infty^{(m_{imp})}(\tau_\alpha)=1 
  ,~~
  Q_\infty^{(1)}(\tau_\alpha)=\frac{e^{-1}-c}{1-c}
  ,
  &\mbox{ if }& c\le1/e,\mbox{ and}
                \nonumber\\
  Q_\infty^{(m_{imp})}(\tau_\alpha)=\frac{1}{ec}
  ,~~
  Q_\infty^{(1)}(\tau_\alpha)=0
  ,~~~~~~
  &\mbox{ if }& c>1/e,
                \nonumber
\end{eqnarray}
\begin{eqnarray}
  \Rightarrow
  \Delta Q_\infty(\tau_\alpha)&=&
                                  \left\{
                                  \begin{matrix}
                                    \frac{1-e^{-1}}{1-c} & \mbox{ if } c\le1/e,&\mbox{ and}\\
                                    \frac{1}{ec} & \mbox{ if } c>1/e.&
                                  \end{matrix}
                                                                       \right.
\end{eqnarray}
\begin{figure}[h]
  \vspace{-.15in}
  \centering
  \includegraphics[width=.9\columnwidth]{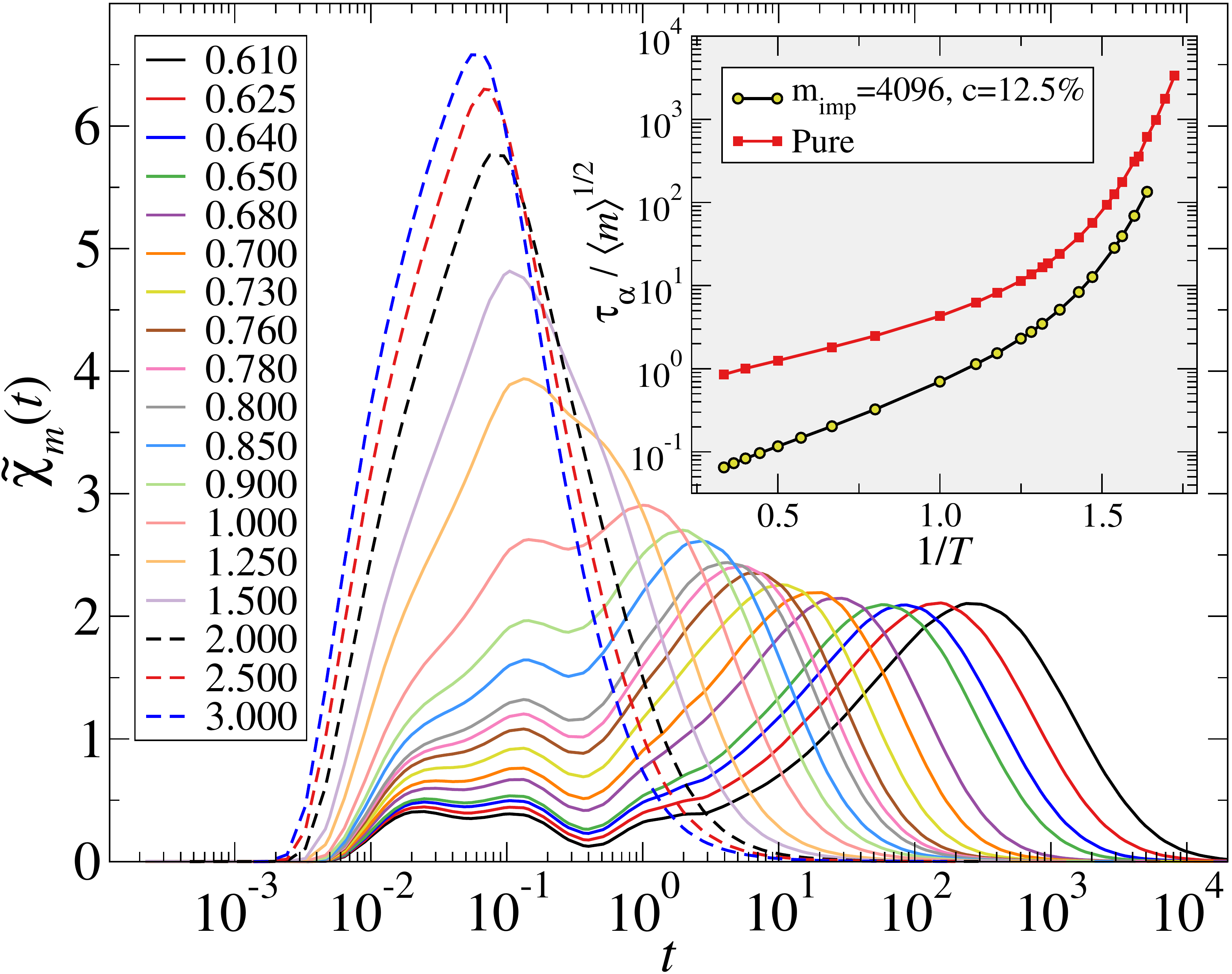}
  \caption{(Color online) 
    $\tilde{\chi}_m(t)$
    versus time for various temperatures in a system with $c=12.5\%$ and
    $m_{imp}=4096$.
    Inset: $\frac{\tau_\alpha}{\sqrt{\langle m\rangle}}$ plotted against $1/T$.
  }
  \label{tauRel}
\end{figure}
\begin{figure*}
  \vspace{-.15in}
  \centering
  \includegraphics[height=0.75\columnwidth]{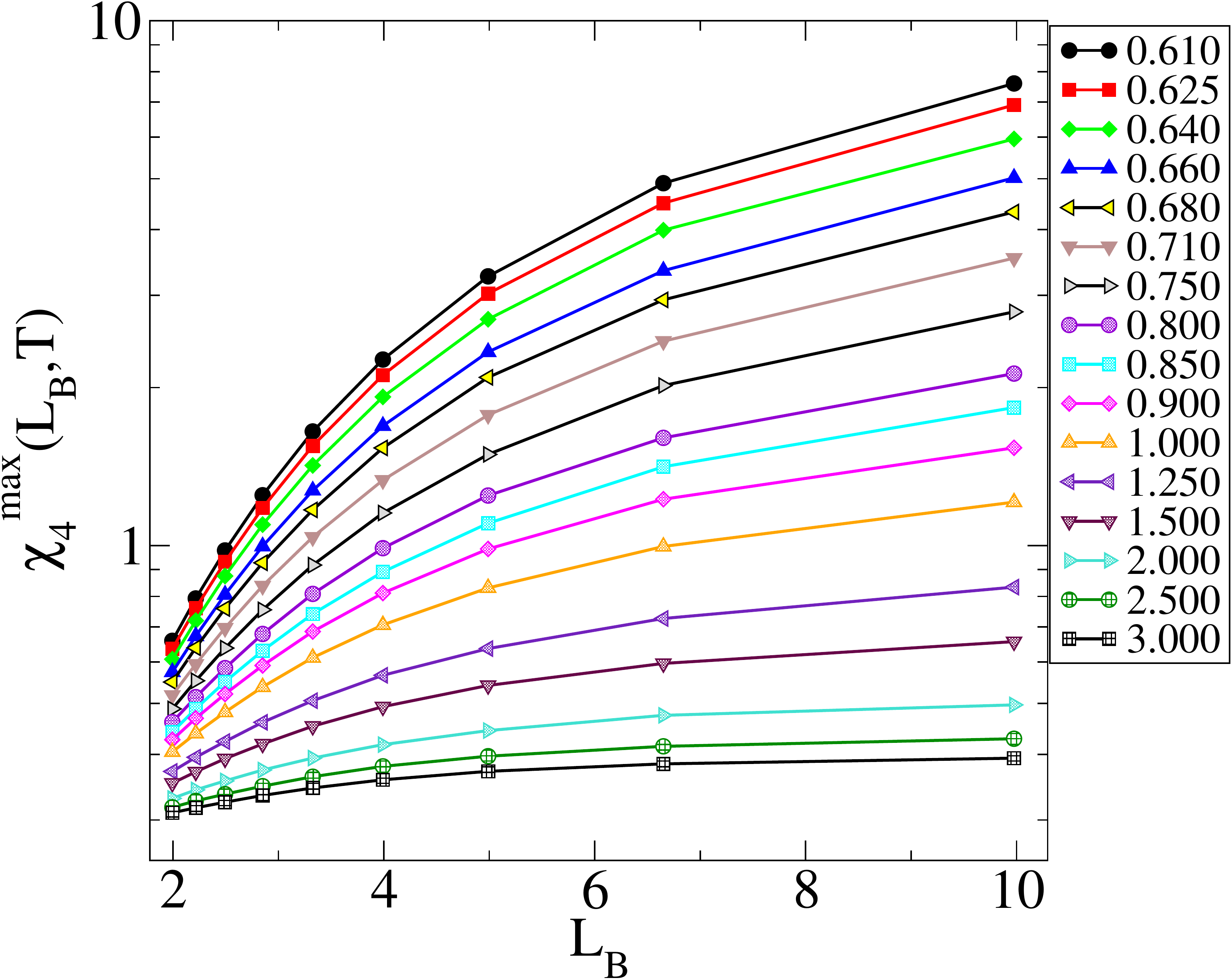}\qquad
  \includegraphics[height=0.75\columnwidth]{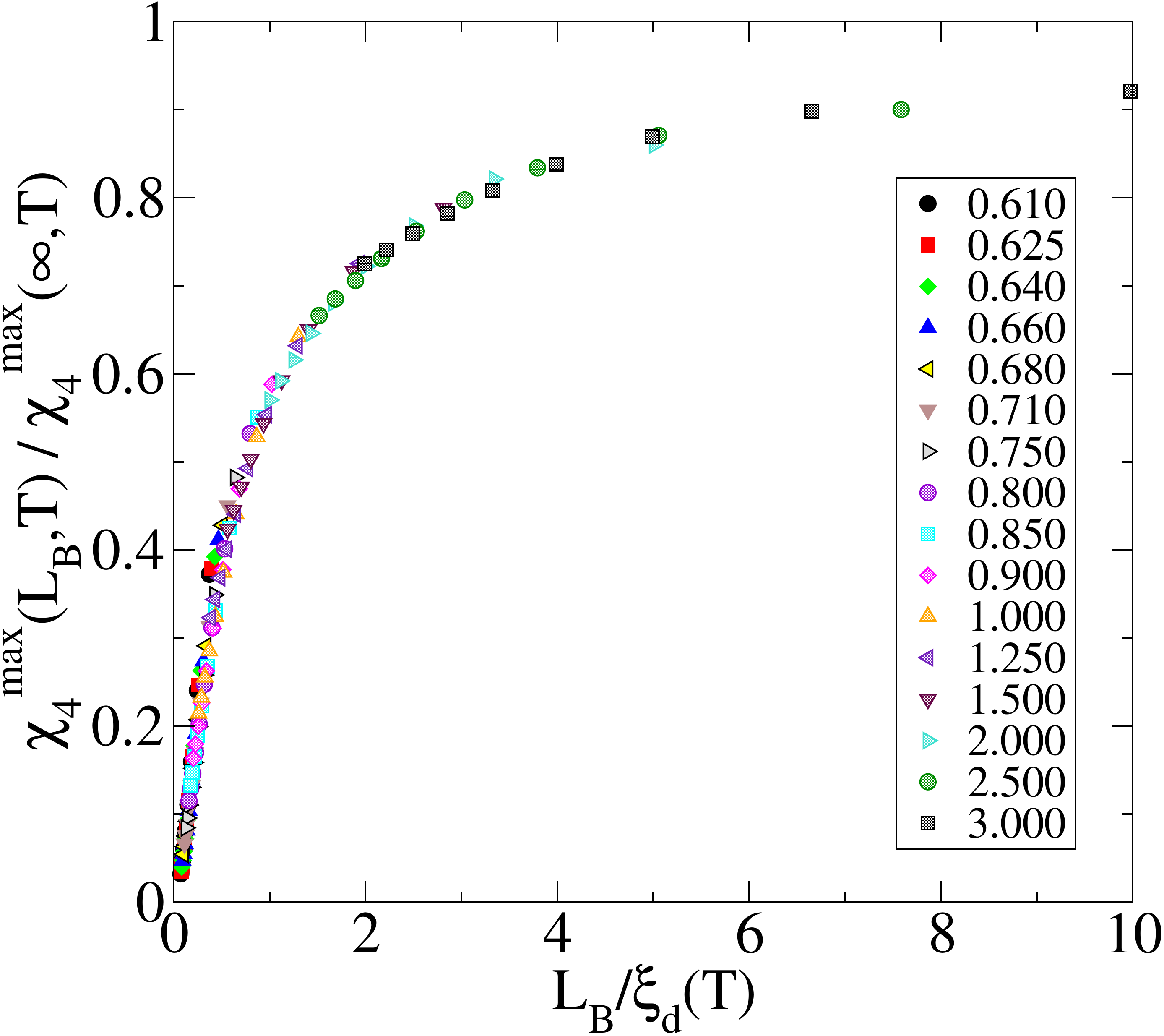}\\
  \includegraphics[height=0.7\columnwidth]{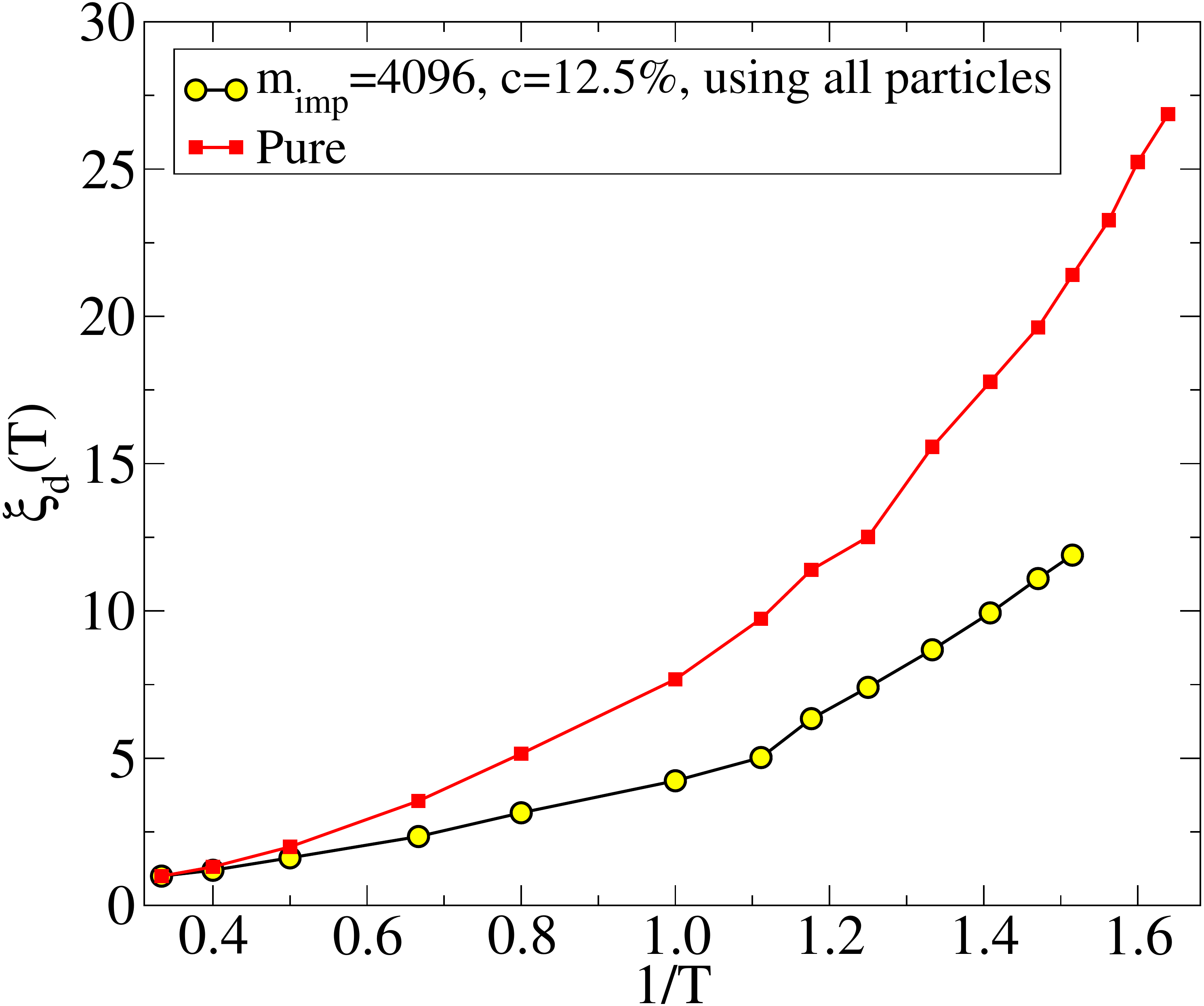}\qquad
  \includegraphics[height=0.7\columnwidth]{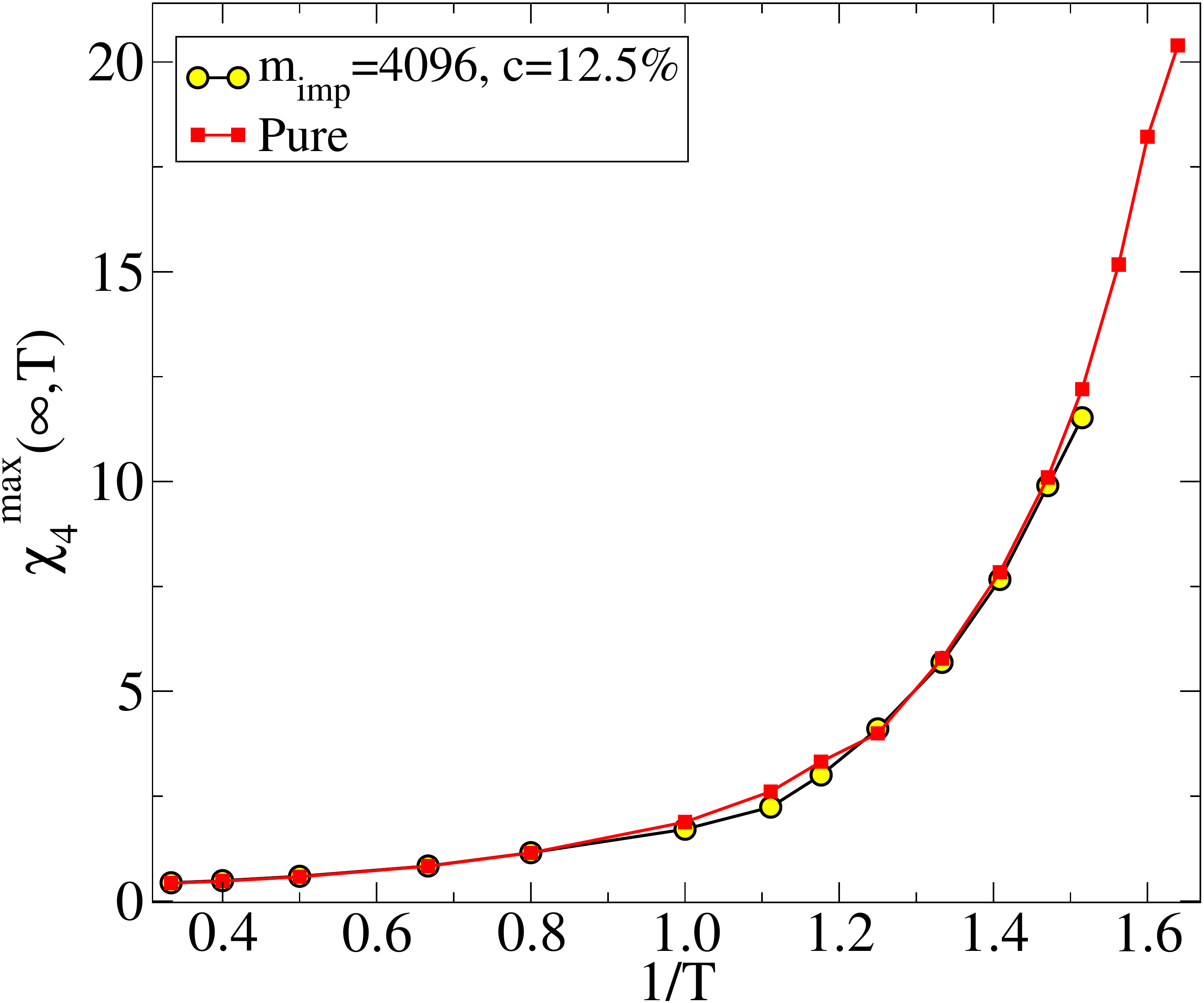}
  \caption{(Color online) 
    Top left: $\chi_4^{\max}$ plotted against block size for various
    temperatures for the pure system.
    Top right: Data collapse using $\chi_4^{\max}$ on the vertical axis
    and $\xi_d(T)$ on the horizontal axis.
    The corresponding plots for the system with impurities look
    qualitatively similar and have therefore not been put here.
    Bottom left: $\xi_d(T)$ versus $1/T$.
    Bottom right: $\chi_4^{\max}(\infty,T)$ versus $1/T$.
  }
  \label{xidFigs}
\end{figure*}

\section{Effect of heavy impurities and effect of mass
  heterogeneity}
\paragraph{Susceptibility to heavy impurities:}
The susceptibility $\chi_m(c,m_{imp},t)$ to heavy impurities can be
defined as 
\begin{eqnarray}
  \chi_m(c,m_{imp},t)=\frac{Q(c,m_{imp},t)-Q(t)}{c}.
  \label{chim}
\end{eqnarray}
The arguments $c$ and $m_{imp}$ have been suppressed in the following
discussion and their values have been held fixed at 12.5\% and 4096,
respectively.
This can be related to the pinning susceptibility, $\chi_p(t)$,
\cite{dck16} via the relation,
\begin{eqnarray}
  \chi_p(t)=\lim_{c\to0}\left[\lim_{m_{imp}\to\infty}\chi_m(c,m_{imp},t)\right].
\end{eqnarray}
We plot $\chi_m(t)$ versus $t$ for varying temperature in
Fig. \ref{chimFig}. 
Unlike the pinning susceptibility $\chi_p$, which keeps growing with
decreasing temperature, the susceptibility to heavy impurities
$\chi_m$ stops growing below a temperature $T^*$ which depends on $c$
and $m_{imp}$. This temperature can be thought of as the temperature
at which the system responds most to the imposed perturbation. It
is interesting that the temperature $T_0$ below which the heavy and
light particles achieve similar levels of relaxation at $\tau_\alpha$
and the temperature $T^*$ at which the dynamics of the system with
impurities is the most different from the pure systems are close to
each other.
\begin{figure}
  \vspace{-.15in}
  \centering
  \includegraphics[height=.7\columnwidth]{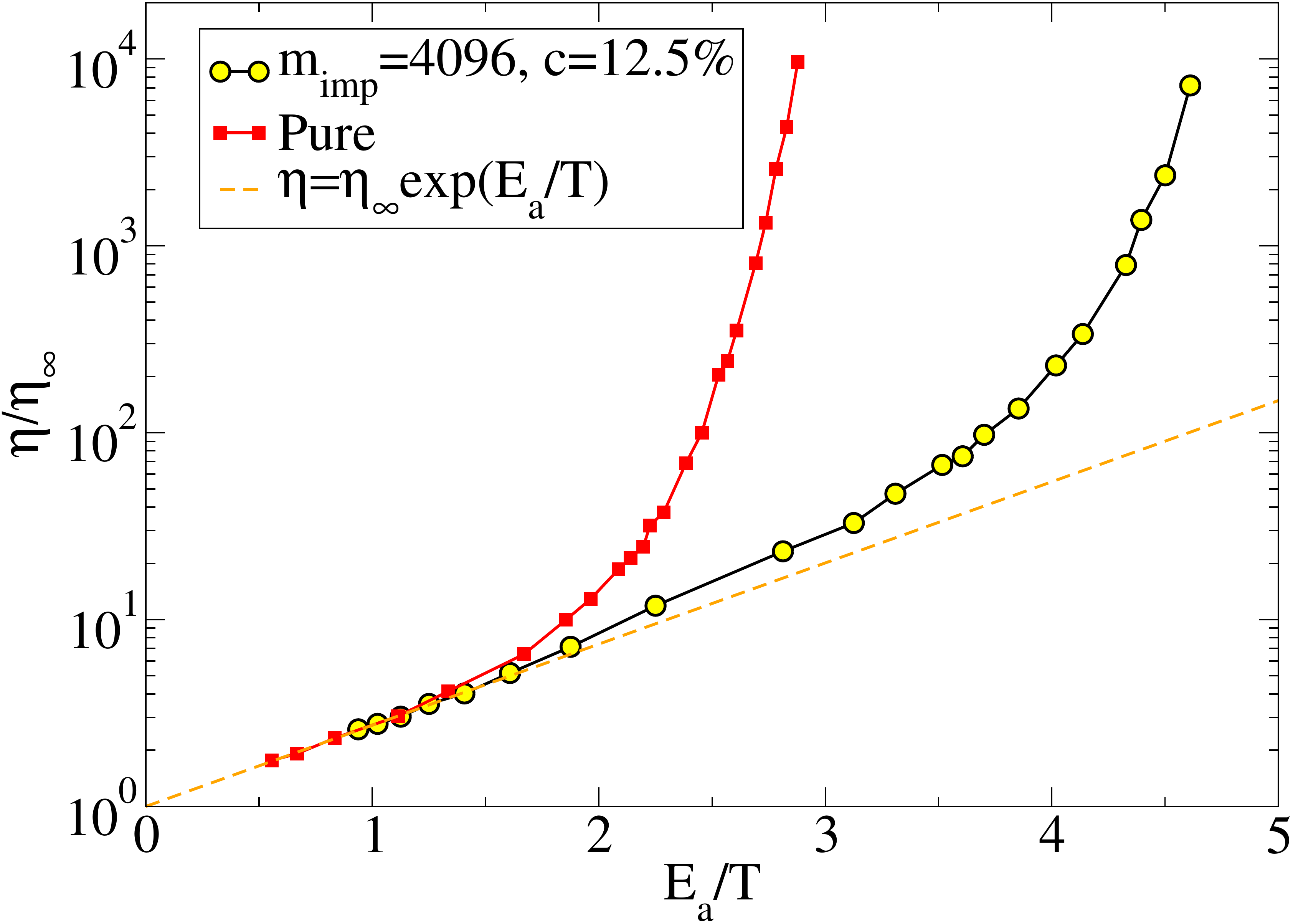}\\ %\qquad
  \includegraphics[height=.7\columnwidth]{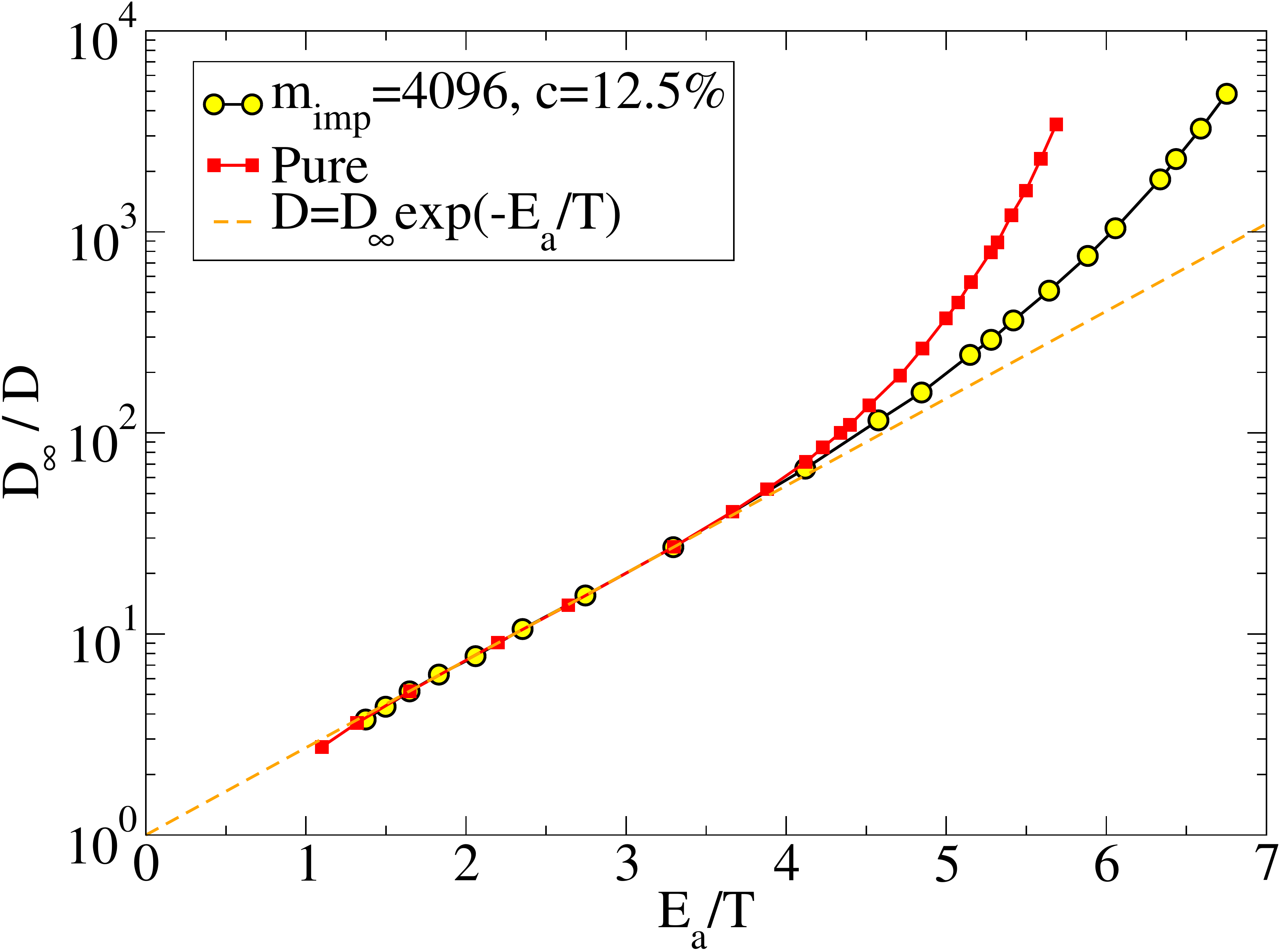}
  \caption{(Color online) 
    Top: Arrhenius plot for the viscosity.
    Bottom: Arrhenius plot for the diffusion constant.
    The constants $\{\eta_\infty, D_\infty, E_a\}$ are fitting
    parameters for the Arrhenius fit 
    (dashed lines) to the high temperature data.
  }
  \label{arrhenius}
\end{figure}

\paragraph{Susceptibility to mass heterogeneity:}
So far in this
work, we have focused on the effect of the heavy impurities. In this
section, we isolate the effect of the heterogeneity in mass from the
overall change in mass. 
We compare dynamics of two systems with the same average mass of the
particles, equal to unity. 
\begin{enumerate}
\item In the first system, all particles have unit mass, $i.e$, this
  is the same as the pure system studied in earlier sections.
\item In the other system, a fraction $c$ of particles are heavier by a
  factor $m_{imp}$, $i.e.$, the lighter particles are of mass
  $\frac{1}{cm_{imp}+1-c}$ and the heavier ones are of mass
  $\frac{m_{imp}}{cm_{imp}+1-c}$.
  The dynamics of this system can be obtained trivially from the system
  with heavy impurities described in earlier sections, just by dividing
  the timescales by the square root of the average mass of particles,
  $\langle m\rangle=cm_{imp}+1-c$.
\end{enumerate}
With this rescaling of time, we find that
the system with the heterogeneity in mass, relaxes
faster than the one with particles of identical mass. 
This is shown in the inset of Fig. \ref{tauRel}.
Similar to the susceptibility to heavy impurities defined in
Eq. \ref{chim}, we can define the susceptibility to mass heterogeneity
quantifying the speedup in relaxation imparted by the mass
heterogeneity.
\begin{eqnarray}
  \tilde{\chi}_m(c,m_{imp},t)=\frac{Q(t)-Q\left(c,m_{imp},t\sqrt{cm_{imp}+1-c}\right)}{c}.
  \nonumber\\
  \label{chimMass}
\end{eqnarray}
This quantity is plotted in Fig. \ref{tauRel}. It is clear that the
presence of mass heterogeneity is felt the most at high temperatures and
its effect diminishes with decreasing temperature.

\section{The dynamic length-scale}
% Dynamical heterogeneity is a phenomenon that is observed in a wide
% variety of supercooled liquids. 
% There is a spatial heterogeneity in
% dynamics characterized by a length-scale $\xi_d(T)$ and this
% length-scale grows as the supercooled liquid is
% cooled. 
The growing spatial heterogeneity in dynamics with decreasing
temperature in a supercooled liquid can be quantified using the dynamic
length-scale $\xi_d(T)$.
This length-scale can be calculated by first
calculating the dynamic susceptibility,
\begin{eqnarray}
  \chi_4(t)=N\langle [Q(t) - \langle Q(t)\rangle]^2\rangle
  \label{chi4defn}
\end{eqnarray}
and then looking at the system size dependence of its
peak value, $\chi_4^{\max}(T)$.
[{\em Note}: In Eq. \ref{chi4defn}, to get the dynamic susceptibility
of the system with impurities, one must plug in $Q(c,m_{imp},t)$
instead of $Q(t)$.] 
In this work, we calculated
the dynamic length-scale using the block analysis technique applied to
$\chi_4^{\max}(T)$.\cite{ctkd17}
In Fig. \ref{xidFigs}, we show the calculation of $\xi_d(T)$ via
$\chi_4^{\max}(T)$.
The top left plot shows the block size dependence of
$\chi_4^{\max}(T)$ for various temperatures for the pure system. This
data is collapsed in the top right plot by rescaling the vertical
and horizontal axes by the thermodynamic limit of $\chi_4^{\max}(T)$
and the dynamic length-scale, respectively. Similar plots for the
system with impurities have not been included here as they
do not appear qualitatively different.
The bottom left and right plots of Fig. \ref{xidFigs} show respectively the
variation of the dynamic length-scale and $\chi_4^{\max}(\infty,T)$
with inverse temperature.
% There are two ways to interpret the
% result. 
% The first (more obvious) way is to think about the growth of
% the dynamic length-scale from its high temperature value.
% To see this clearly,
% we plot $\chi_4^{\max}(\infty,T)$ versus $\xi_d(T)$ after rescaling
% the data to match at a high temperature (chosen as $T=3.00$ here). This is shown in
% the top right panel of Fig. \ref{xidFigs}.
We find that the system with impurities has a slower growth of the
dynamic length-scale with decreasing temperature. Surprisingly
however, the peak value of the dynamic susceptibility in the
thermodynamic limit is not significantly different in the pure system
and the one with impurities.
% \\
% An alternative way is to think about the growth of the dynamic
% length-scale in a temperature range that is deeply in the supercooled
% regime (and observe its extrapolated divergence at the ideal glass
% transition temperature). 
% This is shown in the bottom right panel of Fig. \ref{xidFigs}. 
% In this case the data is matched at $T=0.66$.
% From this, we can clearly see that once the length and time scales are
% sufficiently large, the presence 
% of the heavy impurities have very little effect on the dynamics of the
% supercooled liquid. In particular, the extrapolated divergence of the dynamic
% length and that of the peak value of the dynamic susceptibility still happen
% at the same temperature and in the same way as in the pure system.

\section{How can increase in mass heterogeneity not result in an
  increase in dynamical heterogeneity?}
In this section, we try to explain how a system made out of particles of
varying mass does not always have more dynamical heterogeneity than a
system without any heterogeneity in particle masses. 
Dynamical heterogeneity is the dynamic spatial segregation of
particles on the basis of their mobilities. 
When all particles in the liquid have the same
mass, the system can arbitrarily choose which particles become the
fast particles and which ones become slow at a given instant in
time. Close proximity of fast and slow particles are avoided because
of viscosity, and since the viscosity grows with decreasing temperature,
a growing dynamic length-scale is observed. The presence of heavier
impurity particles imposes additional constraints. Because of their
inherent inertia, the heavier particles cannot change their speeds as
spontaneously as the light particles and therefore have a smaller fluctuation
in their dynamics. It is plausible that starting from a high
temperature at which the dynamics of the pure system is almost
homogeneous, as one lowers the temperature, the growth of the dynamic
clusters are obstructed by the heavy particles resulting in a slower
growth of the dynamic length-scale.
% (as shown in the bottom left plot
% of Fig. \ref{xidFigs}).

\section{Effect on fragility} 
% Even though the main message of this Letter has been theoretical, there
% is an applied perspective. 
% The kinetic fragility of the system with
% impurities can be significantly smaller than that of the pure system
% (depending on the threshold chosen for defining $T_g$). 
% This is clearly seen in the Angell plot shown in Fig. \ref{angell}.
The introduction of heavy impurities can prolong (in terms of
temperature range) the regime of Arrhenius relationship between the
relaxation time or the viscosity and the temperature. In other words,
the kinetic fragility of the system with heavy impurities is 
% significantly 
smaller than that of the pure system.
This is clearly seen in the Arrhenius plots shown in
Fig. \ref{arrhenius}. The calculation of the viscosity was done using
the Green-Kubo relation using the autocorrelation of the off-diagonal
components of the stress tensor.
\begin{eqnarray}
  \eta=\frac{V}{k_BT}\int_0^\infty \langle P_{xz}(0)P_{xz}(t)\rangle
  dt,
\end{eqnarray}
where $P_{xz}$ is an off-diagonal component of the stress tensor,
${\cal P}$, given by,
\begin{eqnarray}
  {\cal P}=\frac{1}{V}\sum_i\left(
  \frac{\vec{p}_i\vec{p}_i}{m_i}+
  \sum_{j>i}\vec{r}_{ij}\vec{F}_{ij}
  \right),
\end{eqnarray}
where the vector products are outer products.
% Even though the plot with the viscosity is qualitatively similar to
% the one with the relaxation time, we have included it here because of
% its experimental appeal. 
A similar result is also present in the
context of random pinning.\cite{CKD15srep} However, making a system with
heavy impurities can be much easier in experiments than randomly
pinning particles. In experiments, both the process of pinning as
well as keeping the pinning random can be quite challenging.
Putting in heavy impurities can be quite straightforward
both in colloidal mixtures consisting of solid and hollow particles
and also in many atomic systems where 
heavier isotopes could be used as the impurity particles.
The thermodynamic fragility at low temperatures however 
becomes indifferent to the presence of the heavy particles since 
at sufficiently low temperatures, the ratio of the
relaxation time of the pure system and the one with impurities veers
towards a constant value.  
Also included in Fig. \ref{arrhenius} are Arrhenius plots for the
% ratio, $\frac{T}{D}$, of the temperature to the corresponding
diffusion constant which also shows a qualitatively similar result.

\section{Conclusions}
In conclusion, we have reported that the presence of heavy impurities
is a method of altering the dynamics of a supercooled liquid without
affecting its static properties. 
Our main finding is the existence of a crossover temperature below
which the heavy and light particles contribute equally to the
relaxation of the system and above which the dominant contribution to
relaxation comes from the lighter particles.
In addition, we have shown suppression of the growth of dynamical
heterogeneity and a decrease in the kinetic fragility as a result of
the introduction of heavy particles in the system, which suggest a way
of using mixture of heavy and light particles for designing new
materials with desired properties. 

\section{Acknowledgements}
This work has been supported in part by the Singapore Ministry of
Education through the Academic Research Fund MOE2019-T2-2-010 and
RG104/17 (S), by Nanyang Technological University Start-Up Grant
(NTU-SUG: M4081781.120), by the Advanced Manufacturing and Engineering
Young Individual Research Grant (A1784C0018) and by the Science and
Engineering Research Council of Agency for Science, Technology and
Research Singapore. We thank NSCC for granting computational
resources.

\section{Data and materials availability}
All data needed to evaluate the conclusions in the paper are presented
in the paper and/or the Supplementary Materials. Additional data
related to this paper may be requested from the authors. 

\begin{appendix}
  
  \section{Dependence of the crossover temperature on $c$ and
    $m_{imp}$}
\begin{figure}[h]
    \centering
    \includegraphics[width=\columnwidth]{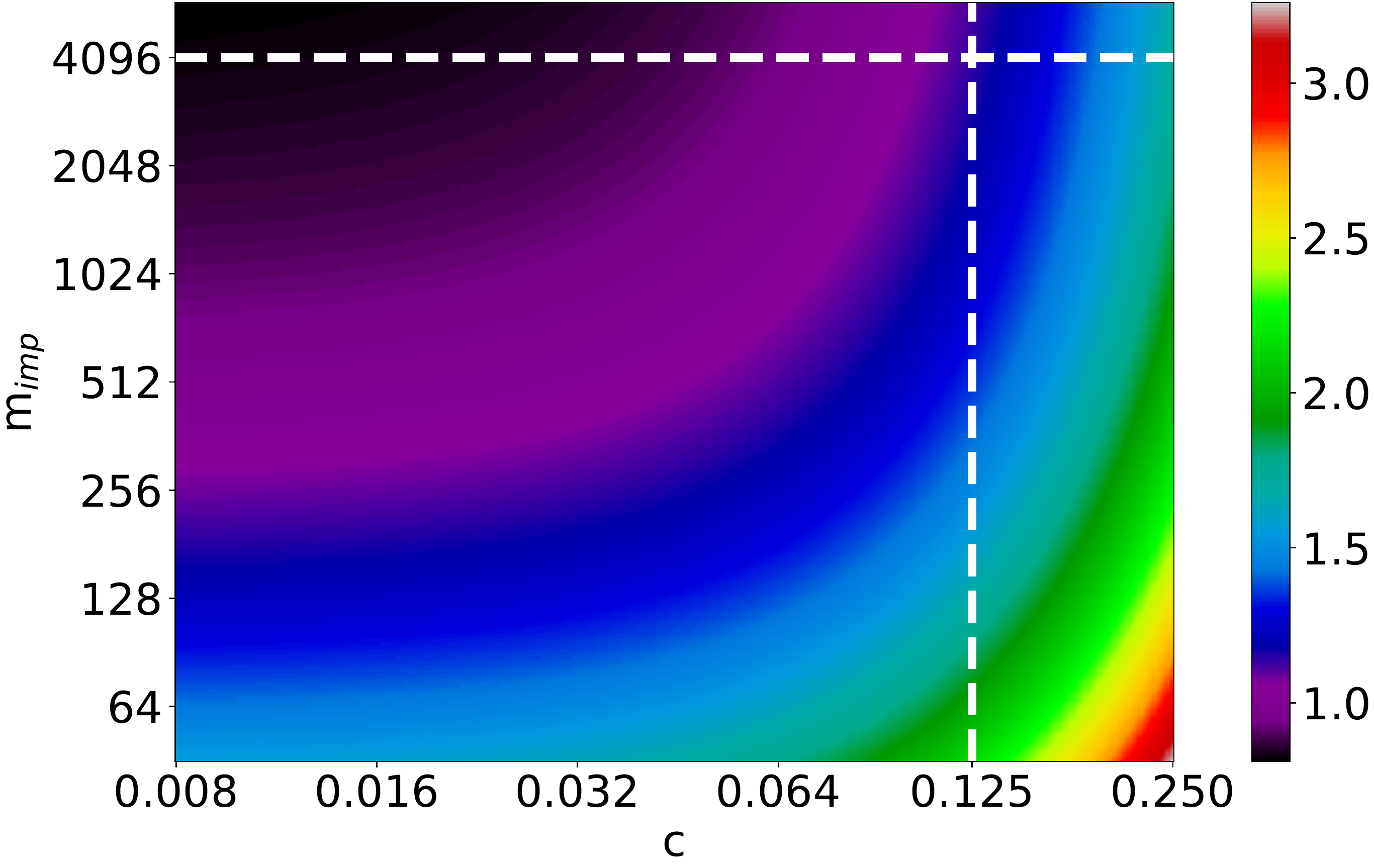}
    \caption{(Color online)
      $T_0$ (in color) as a function of $c$ and $m_{imp}$. The white
      dashed lines correspond to $c=12.5\%$ and $m_{imp}=4096$ which were
      used in the insets of Fig. 1 of the manuscript. 
    }
    \label{T0color}
  \end{figure}

  In this appendix, we investigate the possible existence of a
  relationship of the crossover temperature on $c$ and $m_{imp}$. Even
  though, it is unlikely that a simple relationship $T_0(c,m_{imp})$
  can be obtained from first principles, the trend near points in
  the $c-m_{imp}$ plane where we have data can be looked at.
  We look for such a relationship using a
  simple form for $T_0(c,m_{imp})$ chosen to be of the type
  $f(c)g(m_{imp})$ where $f$ and $g$ are simple functions.
  We know that as $m_{imp}\to1$, the dynamics of the particles would become
  indistinguishable for any temperature. Thus, $T_0\to\infty$ as
  $m_{imp}\to1$. We
  find that our data fits well to the form,
  \begin{eqnarray}
    T_0(c,m_{imp})&=&0.734\left(1+8.82c^{3/2}\right)\left(1+\frac{7.40}{\sqrt{m_{imp}-1}}\right).
                      \nonumber\\
    \label{t0cm}
  \end{eqnarray}
  (The exponents $3/2$ and $1/2$ were chosen to be rational numbers
  close to numbers 
  obtained from initial fits and should not be taken seriously.)
  Variation of $T_0$ with $c$ and $m_{imp}$ is shown in Fig. \ref{T0color}
  using Eq. \ref{t0cm}. 
  We alert the reader that such relationships should 
  only be used for interpolations or small extrapolations to
  understand the trend near the parameter values at which the study
  has been done (the white dashed lines).
  
\end{appendix}

%%%%%%% END %%%%%%%%%%%%%%%
% \bibliography{heavy}
%

%%%%%%%%%%%%%%%%%%%%%%%%%%% 
\end{document}